\documentclass{emulateapj}
\slugcomment{Accepted: January 19, 2011}
\shorttitle{Cosmic Ray Helium Hardening}

\shortauthors{Ohira and Ioka}

\begin{document}

\title{Cosmic Ray Helium Hardening}

\author{Yutaka Ohira and Kunihito Ioka}

\begin{abstract}
Recent observations by CREAM and ATIC-2 experiments suggest that 
(1) the spectrum of cosmic ray (CR) helium is harder than that of CR proton below the knee $10^{15}~{\rm eV}$ and (2) all CR spectra become hard at $\gtrsim 10^{11}~{\rm eV/n}$.
We propose a new picture that higher energy CRs are generated in more helium-rich region to explain the hardening (1) without introducing different sources for CR helium.
The helium to proton ratio at $\sim 100$ TeV exceeds the Big Bang abundance $Y=0.25$ by several times, and the different spectrum is not reproduced within the diffusive shock acceleration theory.
We argue that CRs are produced in the chemically enriched region, such as a superbubble, and the outward-decreasing abundance naturally leads to the hard spectrum of CR helium if CRs escape from the supernova remnant (SNR) shock in an energy-dependent way.
We provide a simple analytical spectrum that also fits well the hardening (2) because of the decreasing Mach number in the hot superbubble with $\sim 10^6$ K.
Our model predicts hard and concave spectra for heavier CR elements.
\end{abstract}


\affil{Theory Center, Institute of Particle and Nuclear Studies, KEK (High Energy Accelerator Research Organization), 1-1 Oho, Tsukuba 305-0801, Japan; ohira@post.kek.jp}
\section{Introduction}
Recently, the Cosmic Ray Energetics And Mass (CREAM) has directly observed the CR compositions with high statistics in the wide energy range up to about $10^{14}~{\rm eV}$. 
Interestingly, CREAM shows $N_{\rm p}(E)\propto E^{-2.66\pm 0.02}$ for CR proton and $N_{\rm He}(E)\propto E^{-2.58\pm 0.02}$ for CR helium in the energy region $2.5\times10^{12}~{\rm eV}$--$2.5\times10^{14}~{\rm eV}$, that is, the spectrum of CR helium is harder than that of CR proton \citep{ahn10}.
Although the difference of the spectral index $\Delta s \approx 0.08$ appears small, the implications are of great importance as shown below.
In addition, the spectral index becomes hard by $\sim 0.12$ for CR proton
and by $\sim 0.16$ for CR helium at $\gtrsim 2\times 10^{11}~{\rm eV/n}$ because the Alpha Magnet Spectrometer (AMS) shows $N_{\rm p}(E)\propto E^{-2.78\pm0.009}$ for the CR proton \citep{alcaraz00a} and $N_{\rm He}(E)\propto E^{-2.74\pm0.01}$ for CR helium \citep{alcaraz00b} in the low-energy range $10^{10}~{\rm eV}$--$10^{11}~{\rm eV}$.
These results have been already obtained by the Advanced Thin Ionization Calorimeter-2
(ATIC-2) \citep{panov09}.

For CR electrons, the {\it Fermi} gamma-ray space telescope has recently observed the spectrum of CR electrons in the wide energy range from $7\times10^9~{\rm eV}$ to $10^{12}~{\rm eV}$ \citep{ackermann10}. 
{\it Fermi} shows that the observed date can be fitted by a power law with spectral index in the interval $3.03 - 3.13$ and the spectral hardening at about $10^{11}~{\rm eV}$,
which may have the same origin as that of the CR nuclei.
\citep[For other models, see, e.g.,][and references therein]{kashiyama10,kawanaka10,ioka10}. 
Note that we do not discuss CR positrons in this letter.

Supernova remnants (SNRs) are thought as the origin of the Galactic CRs. 
The most popular acceleration mechanism at SNRs is the diffusive shock acceleration (DSA) \citep{axford77,krymsky77,bell78,blandford78}.
In fact, {\it Fermi} and {\it AGILE} show that middle-age SNRs interacting with molecular clouds emit gamma-rays \citep[e.g.][]{abdo09,tavani10} and 
the gamma-ray observations support that SNRs produce the bulk of Galactic CRs \citep[e.g.,][]{ohira11,li10}.

According to DSA theory, the spectrum of accelerated particles at a shock does not depend on CR elements, but depends only on the velocity profile of the shock.
Thus, naively, recent CR observations seem to show that the acceleration site of CR helium is different from that of CR proton \citep{biermann10}.
However, in this different site scenario, it should be by chance that the observed ratio of CR helium and proton, $N_{\rm He}/N_{\rm p}$, at $10^9~{\rm eV}$ is similar to the cosmic abundance ($Y=0.25$).
Furthermore, the difference of the spectral index $\Delta s \approx 0.08$ means that $N_{\rm He}/N_{\rm p}$ at $10^{14}~{\rm eV}$ is about 3 times higher than that at $10^9~{\rm eV}$.
This enhancement is amazing since the mean helium abundance in the universe is virtually maintained constant.
The stellar nucleosynthesis never enhances the mean helium abundance by a factor, which is the essential reason that the big bang nucleosynthesis is indispensable for the cosmic helium abundance.
To make the enhancement, we should consider inhomogeneous abundance regions.
We show that this leads to the different spectrum of CR proton and helium when escaping from SNRs.

In this letter, considering the inhomogeneous abundance region, we provide a new explanation about (1) the different spectrum of CR proton and helium, even if CR proton and helium are accelerated simultaneously.
Our idea uses the fact that CRs escaping from SNRs generally have a different spectrum than
that of the acceleration site \citep{ptuskin05,ohira10,caprioli10}.
The runaway CR spectrum depends on not only the acceleration spectrum at shocks but also the evolution of the maximum energy and the number of accelerated CRs \citep{ohira10}. We also suggest that (2) the spectral hardening of CRs is caused by the decreasing Mach number in the high temperature medium.
Both the inhomogeneous abundance and the high temperature can be realized in the superbubbles with multiple supernovae.
Our conclusions are summarized as follows.
\begin{itemize}
\item Runway CR spectra depend on not only CR spectra inside the SNR but also the evolution of the maximum energy and the number of accelerated CRs.
Therefore, taking account of the inhomogeneous abundance region, runaway CR spectra of different CR elements have different spectra (section 2 and 3.1).
\item Our model is in excellent agreement with observed spectra of CR proton and helium.
Harder spectrum of CR helium is due to the enhancement of the helium abundance around the explosion center.
On the other hand, the concave spectra of all CR elements are due to the decreasing Mach number in the hot gas with $\sim 10^6~{\rm K}$.
The concave spectra may be also produced by the CR nonlinear effect, the energy dependent effects on the accelerated CRs (on $\alpha$ or $\beta$), the propagation effect ($\gamma$), 
and/or multi components with different spectral indices (section 3.2 and 4).
\item Within the single component scenario, the hard helium spectrum suggests that the origin of the Galactic CR is SNRs in superbubbles, although we are not excluding the multi component scenario (section 5).
\item Our model predicts that heavier (at least volatile) CR elements also have harder spectra than that of CR proton and have concave spectra (section 5).
\end{itemize} 

\section{Runaway CR spectrum}
\label{sec:2}
\begin{figure}
\plotone{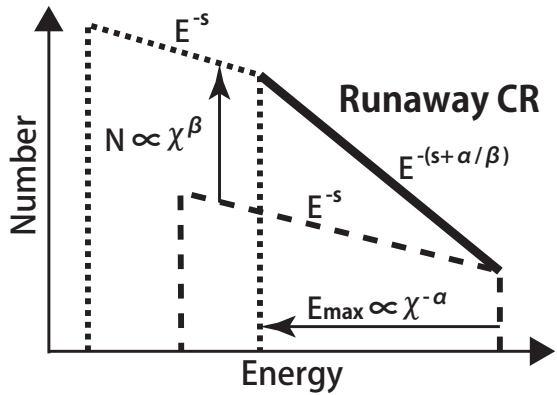} 
\caption{Schematic picture of the runaway CR spectrum. 
The solid, dashed and dotted lines show the runaway CR spectrum, CR spectrum inside an SNR at an early epoch and CR spectrum inside the SNR at a later epoch, respectively.
The solid line of the runaway CR spectrum represents the equation (\ref{eq:Fesc}).
A variable $\chi$ (e.g., the shock radius) describes the SNR evolution.
\label{fig1}}
\end{figure}

In this section, we briefly review the runaway CR spectrum (see Appendix of \citet{ohira10}).
We here use a variable $\chi$ (for example the shock radius or the SNR age) to describe the evolution of an SNR.
Let $F_{\rm SNR}(\chi,p)$ and $p_{\rm max}(\chi)$ be the CR momentum spectrum [(${\rm eV}/c)^{-1}$] and the maximum four momentum of CR inside the SNR at a certain epoch labeled by $\chi$, respectively.
CRs escape in order, from the maximum energy CR because the diffusion length of high-energy CRs is larger than that of low-energy CRs.
Then, the number of runaway CRs between $\chi$ and $\chi+{\rm d}\chi$ is
\begin{equation}
F_{\rm SNR}(\chi,p_{\rm max}) \frac{{\rm d}p_{\rm max}}{{\rm d}\chi} {\rm d}\chi~~,
\end{equation}
which corresponds to the number of runaway CRs between $p=p_{\rm max}(\chi)$ and $p=p_{\rm max}(\chi)+{\rm d}p$, $F_{\rm esc}(p){\rm d}p$. 
Hence, $F_{\rm esc}(p)$ is
\begin{equation}
F_{\rm esc}(p) = F_{\rm SNR}(p_{\rm max}^{-1}(p),p)~~,
\label{eq:esc}
\end{equation}
where $p_{\rm max}^{-1}(p)$ is the inverse function of $p_{\rm max}(\chi)$. 
Assuming $F_{\rm SNR}(\chi,p)\propto \chi^{\beta}p^{-s}$ and $p_{\rm max}(\chi)\propto \chi^{-\alpha}$, we obtain the runaway CR spectrum as
\begin{equation}
F_{\rm esc}(p) \propto p^{-\left(s+\frac{\beta}{\alpha}\right)}~~,
\label{eq:Fesc}
\end{equation}
where $\alpha$ and $\beta$ are parameters to describe the evolution of maximum energy and the number of accelerated CRs, respectively. 
(We use {$\alpha \sim 6.5$} and $\beta \sim 1.5$ later.)
Therefore, the runaway CR spectrum $F_{\rm esc}$ is different from that in the SNR $F_{\rm SNR} \propto p^{-s}$.
Figure~\ref{fig1} shows the schematic picture of the runaway CR spectrum.
In this Letter, we use the shock radius, $R_{\rm sh}$, as $\chi$.

The evolution of the maximum energy of CRs at the SNR has not been understood.
This  strongly depends on the evolution of the magnetic field around the shock \citep[e.g.][]{ptuskin03}.
Although some magnetic field amplifications have been proposed \citep[e.g.,][]{lucek00,bell04,giacalone07,ohira09b} and investigated by simulations \citep[e.g.,][]{niemiec08,riquelme09,ohira09a,gargate10}, the evolution of the magnetic field has not been completely understood yet. 
Here we assume that CRs with the knee energy escape at $R=R_{\rm Sedov}$, where $R_{\rm Sedov}$ is the shock radius at the beginning of the Sedov phase.
Furthermore, we use the phenomenological approach with the power-law dependence \citep{gabici09,ohira10},
\begin{equation}
p_{\rm max}(R_{\rm sh}) = p_{\rm knee} Z \left (\frac{R_{\rm sh}}{R_{\rm Sedov}}\right)^{-\alpha}~~,
\label{eq:pmax}
\end{equation}
where $p_{\rm knee}=10^{15.5}~{\rm eV}/c$ is the four momentum of the knee energy.
Note that $\alpha$ does not depend on the CR composition because the evolution of the maximum energy depends only on the evolution of the magnetic field and the shock velocity.

The evolution of the number of CRs inside the SNR has not been also understood.
This depends on the injection mechanism \citep{ohira10} and the density profile around the SNR.
We here adopt the thermal leakage model \citep{malkov95} as an injection model.
For the total density profile, $\rho_{\rm tot}(R_{\rm sh})\approx m_{\rm p}\left(n_{\rm p}(R_{\rm sh})+4n_{\rm He}(R_{\rm sh})\right)$ where $n_{\rm p}$ and $n_{\rm He}$ are the number density of proton and helium and $m_{\rm p}$ is the proton mass, the shock velocity of the Sedov phase is 
\begin{equation}
u_{\rm sh}(R_{\rm sh}) \propto \rho_{\rm tot}(R_{\rm sh})^{-\frac{1}{2}} R_{\rm sh}^{-\frac{3}{2}}~~.
\label{eq:ush}
\end{equation}
In the thermal leakage model, the injection momentum of element $i$ is proportional to the shock velocity, $p_{{\rm inj},i}\propto u_{\rm sh}$, and the number density of CR with momentum $p_{{\rm inj},i}$ is proportional to the density, $p_{{\rm inj},i}^3f_i(p_{{\rm inj},i})\propto n_i(R_{\rm sh})$, where $f_i$ is the distribution function of CR element $i$.
Hence, the number of CR element $i$ with a reference momentum $p=m_{\rm p}c$, $F_{{\rm SNR},i}(R_{\rm sh},m_{\rm p}c)$ is
\begin{eqnarray}
F_{{\rm SNR},i}(R_{\rm sh},m_{\rm p}c)&\propto& R_{\rm sh}^3f_i(m_{\rm p}c) \nonumber \\
&\propto& R_{\rm sh}^3p_{{\rm inj},i}^{s_{\rm low}+2}f_i(p_{{\rm inj},i}) \nonumber \\
&\propto& R_{\rm sh}^3 n_i(R_{\rm sh})p_{{\rm inj},i}^{s_{\rm low}-1} \nonumber \\
&\propto& n_i(R_{\rm sh})\rho_{\rm tot}(R_{\rm sh})^{\frac{1-s_{\rm low}}{2}}R_{\rm sh}^{\frac{3(3-s_{\rm low})}{2}}~~,
\label{eq:beta}
\end{eqnarray}
where $f_i(p)p^2\propto p^{-s_{\rm low}}$ and $s_{\rm low}$ is the spectral index in the non-relativistic energy region.
For the nonlinear DSA, the spectral index in the non-relativistic energy region is different from that in the relativistic energy region \citep{berezhko99}.
To understand the essential feature of the runaway CR spectrum, we here consider only the test-particle DSA, that is, $s_{\rm low} = s$.
Because $n_i(R_{\rm sh})\rho_{\rm tot}(R_{\rm sh})^{\frac{1-s}{2}}$ is not always a single power-law form, the evolution of the number of accelerated CRs can not be always described by a constant $\beta$.

\section{Basic idea}
\label{sec:3}
\subsection{Different spectrum of CR proton and helium}
\label{sec:3.1}
\begin{figure}
\plotone{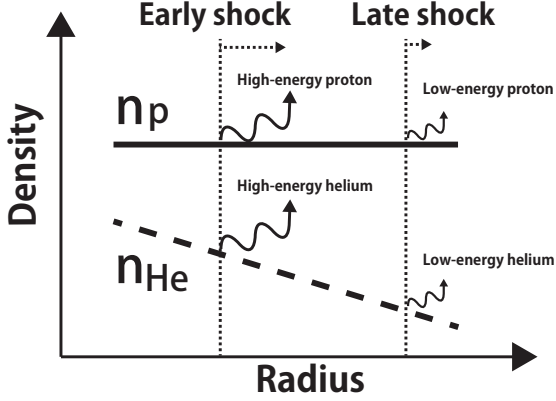} 
\caption{Schematic picture of the formation of the different spectrum.
The solid and dashed line show the proton density and the helium density, respectively.
The dotted lines show the shock front.
In early phase, high-energy CR proton and CR helium escape, and in late phase, low-energy CR proton and CR helium escape.
The ratio of CR helium to CR proton increases with the CR energy.   
\label{fig2}}
\end{figure}
According to the test particle DSA theory, the index $s$ of relativistic CR energy spectrum depends only on the velocity jump at the shock,
\begin{equation}
s = \frac{u_1 +2u_2}{u_1-u_2} = 2\frac{M^{2}+1}{M^{2}-1}~~,
\label{eq:s}
\end{equation}
where we use the Rankine-Hugoniot relation at the second equation and $M$ is the Mach number.
Then, the index of the runaway CR spectrum, $s_{\rm esc}$, is
\begin{equation}
s_{\rm esc} = s + \frac{\beta}{\alpha}~~,
\label{eq:sesc}
\end{equation}
in equation (\ref{eq:Fesc}).
Therefore, if $\beta/\alpha$ (in particular $\beta$, the index for the 
accelerated CR number evolution) is different, the runaway CR spectrum is different between the CR compositions.
This is our main idea to explain the helium hardening observed by CREAM and ATIC-2.
From equation (\ref{eq:beta}), $\beta$ depends on the ambient number density $n_i$.
Therefore, different density profiles make different runaway CR spectra (See Section \ref{sec:4} for more details).
Figure~\ref{fig2} shows the schematic picture of our idea.

\subsection{Spectral hardening of all CRs at the same energy per nucleon}
\label{sec:3.2}
In this subsection, we discuss the spectral hardening of the observed CRs.
The Galactic CR spectrum observed at the Earth, $F_{\rm obs}$, is obtained by the simple leaky box model
\begin{equation}
F_{\rm obs} \propto F_{\rm esc}(p)/D(p) \propto F_{\rm esc}(p) p^{-\gamma} ~~,
\label{eq:Fobs}
\end{equation}
where $D(p)\propto p^{\gamma}$ is the diffusion coefficient \citep[e.g.][]{strong07}.
Hence, the index of the observed spectrum is
\begin{equation}
s_{\rm obs} = s + \frac{\beta}{\alpha}+\gamma~~.
\label{eq:sesc}
\end{equation}
The deviation from a single power law means that at least one of $s$, $\alpha$, $\beta$, and $\gamma$ has an energy dependence or that the origin of low energy CRs below $10^{11}~{\rm eV}$ is different from that of high energy CRs above $10^{11}~{\rm eV}$.
Although the multi component scenario may be the case because there are many types of SNRs, we discuss the single component scenario in this letter.

Firstly, we discuss the energy dependence of $s$.
From equation (\ref{eq:s}), $s$ depends on the shock radius because the Mach number $M$ decreases with the shock radius.
Then we can expect the spectral harding of all CR compositions at the same rigidity $cp/Ze$, that is, at approximately the same energy per nucleon.
From equation (\ref{eq:ush}), the Mach number is
\begin{equation}
M \approx 10^3 \left(\frac{\rho_{\rm tot}(R_{\rm sh})}{\rho_{\rm tot}(R_{\rm Sedov})}\right)^{-\frac{1}{2}}\left(\frac{T}{10^4~{\rm K}}\right)^{-\frac{1}{2}} \left(\frac{R_{\rm sh}}{R_{\rm Sedov}}\right)^{-\frac{3}{2}} ~~,
\label{eq:m}
\end{equation}
where $T$ is the surrounding temperature and we assume that the ejecta mass and the energy of supernova explosion are $1~{\rm M}_{\odot}$ and $10^{51}~{\rm erg}$, respectively.
From equations~(\ref{eq:pmax}), (\ref{eq:s}) and (\ref{eq:m}), we can obtain $s$ as a function of $p$ (see \S~\ref{sec:4}).

Alternatively the spectral hardening can be also interpreted as the CR nonlinear effect \citep[e.g.,][]{drury81,malkov01}.
This issue will be addressed in the future work.

Next, we discuss the energy dependence of $\beta$ that is the parameter to describe the evolution of the number of accelerated CRs.
In Section~\ref{sec:3.1}, we consider different power-law forms for $n_{\rm p}(R_{\rm sh})$ and $n_{\rm He}(R_{\rm sh})$ to make the different spectrum of the CR proton and helium.
Therefore, $\rho_{\rm tot}(R_{\rm sh}) \approx m_p [n_p(R_{\rm sh})+4 n_{\rm He}(R_{\rm sh})]$ is not a single power law form, and $\beta$ has an energy dependence (see \S~\ref{sec:4}).

The energy dependence of $\gamma$ will be soon precisely determined by AMS-02 \citep{pato10}.
We do not discuss the energy dependence of $\alpha$ 
because the complete physics of the CR escape and magnetic turbulence 
is beyond the scope of this Letter.

\section{Comparison of our model with observations}
\label{sec:4}

\begin{figure}
\plotone{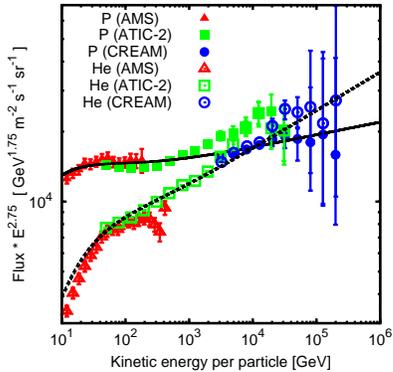} 
\caption{Comparison of our model (solid and dashed line) with AMS (triangle) \citep{alcaraz00a,alcaraz00b}, ATIC-2 (square) \citep{panov09} and CREAM (circle) \citep{ahn10} observations for Galactic CRs.
Filled symbols and the solid line show CR proton.
Open symbols and the dashed line show CR helium.
\label{fig3}}
\end{figure}
\begin{figure}
\plotone{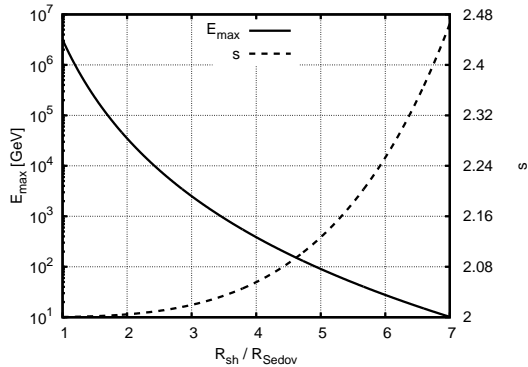} 
\caption{The evolution of the maximum energy $E_{\rm max}/Z$ (solid line) and the spectral index $s$ of test particle DSA (dashed line) in equations (\ref{eq:pmax}), (\ref{eq:s}) and (\ref{eq:m})
as functions of $R_{\rm sh}$.
\label{fig4}}
\end{figure}
In this section, specifying model parameters, we calculate the Galactic CR spectrum.
For simplicity, we here assume the number densities of proton and helium as follows,
\begin{eqnarray}
n_{\rm p}(R_{\rm sh})&=&n_{\rm p,0}\nonumber \\
n_{\rm He}(R_{\rm sh})&=& \zeta n_{\rm p,0} \left(\frac{R_{\rm sh}}{R_{\rm Sedov}}\right)^{-\delta}~~,
\label{eq:npnHe}
\end{eqnarray}
where $n_{\rm p,0}$ is the number density of proton at $R_{\rm sh} = R_{\rm Sedov}$, and $\zeta n_{\rm p,0}$ is the normalization factor of the helium density. 
We set $\zeta=10^{6.5 (\delta/\alpha)-1}$ so that the helium abundance is that of the solar abundance, $n_{\rm He}/n_{\rm p}=0.1$
 (i.e., $Y \approx 0.25$), when $cp_{\rm max}=Z~{\rm GeV}$ with equation (\ref{eq:pmax}).
Note that the power-law dependence is a first step approximation for the mean value.
Then, from equations~(\ref{eq:esc}), (\ref{eq:pmax}), (\ref{eq:beta}), (\ref{eq:Fobs}), observed spectra of CR proton and helium are
\begin{eqnarray}
F_{\rm obs,p} &=& F_{\rm p,knee}\left\{\frac{1+\zeta\left({p}/{p_{\rm knee}}\right)^{\frac{\delta}{\alpha}}}{1+\zeta}\right\}^{\frac{1-s(p)}{2}} \nonumber \\
&&~~~~~~~~~~~~~\times \left(\frac{p}{p_{\rm knee}}\right)^{-\left[s(p)+\frac{3\left\{3-s(p)\right\}}{2\alpha}+\gamma\right]},\\
F_{\rm obs,He} &=& \epsilon F_{\rm p,knee}\left\{\frac{1+\zeta\left({p}/{Zp_{\rm knee}}\right)^{\frac{\delta}{\alpha}}}{1+\zeta}\right\}^{\frac{1-s(p)}{2}} \nonumber \\
&&~~~~~~~~~~~~~\times\left(\frac{p}{Zp_{\rm knee}}\right)^{-\left[s(p)+\frac{3\left\{3-s(p)\right\}-2\delta}{2\alpha}+\gamma\right]}~~,
\end{eqnarray}
where $F_{\rm p,knee}$ and $\epsilon F_{\rm p,knee}$ are normalization factors of CR proton and helium and $Z=2$ for helium, and $s(p)$ is obtained from equations (\ref{eq:pmax}), (\ref{eq:s}) and (\ref{eq:m}).
In this model, all parameters are $\alpha$, $\gamma$, $\delta$, $\epsilon$, $T$, $F_{\rm p,knee}$.

Figure~\ref{fig3} shows the comparison of our model with observations.
We take into account the solar modulation effects with the modulation potential $\Phi = 450~{\rm MV}$ \citep{gleeson68}.
Our model is in excellent agreement with the observed spectra, with $\alpha=6.5$, $\gamma=0.43$, $\delta=0.715$, $\epsilon=0.31$, $T=10^6~{\rm K}$.
The different spectra of CR proton and helium originate from the different density profiles in equations (\ref{eq:npnHe}).
Figure~\ref{fig4} shows the evolution of the maximum energy of CRs and the spectral index of CRs inside the SNR. 
In the early phase, the spectral index $s$ is 2 and after then, the spectral index decreases with the shock radius because the Mach number decreases with shock radius. 
The change of spectral index $s$ is about 0.1 which is almost the same as the observed hardening.
The observed hardening is not the result of the change of the injection history, $\beta$, but the result of spectral change of CRs inside the SNR.
The high temperature $T\sim10^6~\rm{K}$ is necessary for the spectral hardening $\Delta s \sim 0.1$.

In addition, our model also makes a concave spectrum of CR electrons as observed
\citep{ackermann10}. 
However, the evolution of injection efficiency of CR electrons has not been understood well. 
So we need further studies to discuss the CR electron spectrum in detail. 

\section{Discussion}
\label{sec:5}
To make the different spectrum, our model requires that the helium abundance around the explosion center is higher than that of the solar abundance.
SNRs in superbubbles are one of candidates.
\citet{higdon98} show that supernova ejecta can dominate the superbubble mass within a core radius of one third of the superbubble radius. 
In the stellar wind and the supernova explosion, the stellar hydrogen envelope has lower density and higher velocity than that of helium.
Then we expect that the helium fraction in the center of superbubbles is higher than that in the outer region.
Furthermore, to make the concave spectrum, our model requires an ambient medium with high temperature, $T=10^6~{\rm K}$.
This is also consistent with superbubbles.
According to the CR composition study, SNRs in superbubbles have been considered as the origin of Galactic CRs \citep[e.g.,][]{lingenfelter07,ogliore09}. 
Particle accelerations in superbubbles have been also investigated by intensive studies \citep[e.g.,][]{bykov92,parizot04,dar08,ferrand10}. 
We here considered a spherically symmetric system. 
The off-center effects may be important for the initial  phase and thereby for the high energy spectrum, because the shock radius at the beginning of the Sedov phase $R_{\rm Sedov}$ is about $20~{\rm pc}$ which is comparable to the typical size of OB association, $35~{\rm pc}$ \citep{parizot04}, and the shock radius $R_{\rm sh}$ is about $200~{\rm pc}$ at the end of the Sedov phase. 
This is an interesting future problem.

Note that the spectral hardening can be also made by the nonlinear model, the energy dependence of the CR diffusion coefficient and/or multi components with different spectral indices. 
So the high temperature may not be absolutely necessary. 
The stellar wind of red giants is one of candidates for the cold and helium rich ambient.
Still, the dominant core-collapse supernovae is type II \citep[e.g.,][]{smartt09} which has no helium rich wind, so that the superbubble scenario looks more likely as the origin of the Galactic CRs above $10^{11}~{\rm eV}$.
For the CRs below $10^{11}~{\rm eV}$, the spectral difference between CR proton and helium may be caused by the solar modulation and the inelastic interactions \citep{putze10}.

The spatial variation of the helium ionization degree can also change the injection history.
The injection efficiency of the large rigidity is thought to be higher than that of low rigidity since particles with large rigidity can easily penetrate through the shock front from the downstream region.
If the ionization degree increases with the SNR radius,
the CR helium spectrum becomes harder than the CR proton one,
$\beta_{\rm He}<\beta_{\rm p}$.
However, the rigidity dependence of the injection efficiency has not been understood completely.
Moreover, the injection from neutral particles should also be understood \citep{ohira09b,ohira10}.

According to our model, CR spectra of heavier volatile elements than helium is also harder than that of proton.
Low-energy CRs of refractory elements are thought to result from suprathermal injection by sputtering off preaccelerated, high-velocity grains \citep{ellison97}.
To be accelerated to the relativistic energy, the refractory elements should be sputtered because the grains can not be accelerated to the relativistic energy.
The SNR shock velocity is not fast enough to accelerate refractory elements to the knee energy when the refractory elements are injected because the sputtering time scale is too long.
Therefore, refractory CRs around the knee energy should be injected by the standard manner similar to volatile CRs.
In this case, the refractory CRs also have harder spectra than protons, although we need further studies of the injection of refractory CRs at the knee energy.

If CRs trapped inside the SNR and released at the end of the SNR's life outnumber runaway CRs
(see figure 3 in \citet{caprioli10}), our scenario does not work for producing hard and concave spectra.
In our model with $\alpha\sim 6.5$ in Eq.~(\ref{eq:pmax}), trapped CRs have energy below $1~{\rm GeV}$ when they are released, that is, $p_{\rm max}\lesssim Zm_{\rm p}c$ when $R_{\rm sh}\gtrsim10R_{\rm Sedov}$, and are not relevant for our interest.
Higher energy CRs escape from the SNR even after advected to the downstream  since the CR diffusion is faster than the expansion of the SNR.
Our case is similar to the right figure 7 in \citet{caprioli10} where trapped CRs are released below $100~{\rm GeV}$.
The energy boundary between trapped CRs and runaway CRs depends on the evolution of the maximum energy ($\alpha$).

\acknowledgments
We thank the referee, T. Suzuki, T. Terasawa and A. Bamba for comments.
This work is supported in part by grant-in-aid from the Ministry of Education, Culture, Sports, Science, and Technology (MEXT) of Japan, No.~21684014 (Y.~O. and K.~I.), Nos.~19047004, 22244019, 22244030 (K.~I.).


\begin{thebibliography}{}
%
\bibitem[Abdo et al.(2009)]{abdo09} Abdo, A. A., et al., 2009, \apj, 706, L1
%
\bibitem[Ackermann et al.(2010)]{ackermann10} Ackermann, M., et al., 2010, \prd, 82, 092004
%
\bibitem[Ahn et al.(2010)]{ahn10} Ahn, H. S., et al., 2010, \apj, 714, L89
%
\bibitem[Alcaraz et al.(2000a)]{alcaraz00a} Alcaraz, J., et al., 2000a, Phys. Lett. B, 490, 27
%
\bibitem[Alcaraz et al.(2000b)]{alcaraz00b} Alcaraz, J., et al., 2000b, Phys. Lett. B, 494, 193
%
\bibitem[Axford et al.(1977)]{axford77}
Axford, W. I., Leer, E., \& Skadron, G., 1977, Proc. 15th Int. Cosmic Ray Conf., Plovdiv, 11, 132
%
\bibitem[Bell(1978)]{bell78}Bell, A. R. 1978, \mnras, 182, 147
%
\bibitem[Bell(2004)]{bell04} Bell, A. R., 2004, \mnras, 353, 550
%
\bibitem[Berezhko \& Ellison(1999)]{berezhko99}Berezhko, E. G., \& Ellison, D. C., 1999, \apj, 526, 385
%
\bibitem[Biermann et al.(2010)]{biermann10}Biermann, P. L., Becker , J. K., Dreyer, J., Meli, A., Seo, E., \& Stanev, T., 2010, \apj, 725, 184
%
\bibitem[Blandford \& Ostriker(1978)]{blandford78}Blandford, R. D., \& Ostriker, J. P., 1978, \apj, 221, L29
%
\bibitem[Bykov \& Fleishman(1992)]{bykov92}Bykov, A. M., \& Fleishman, D. G., 1992, \mnras, 255, 269
%
\bibitem[Caprioli et al.(2010)]{caprioli10}Caprioli, D., Amato, E., \& Blasi, P., 2010, Astropart. Phys., 33, 160
%
\bibitem[Dar \& De R\'{u}jula (2008)]{dar08}Dar, A., \& De R\'{u}jula, A., 2008, \physrep, 466, 179
%
\bibitem[Drury \& V\"{o}lk(1981)]{drury81}Drury, L. O'C., \& V\"{o}lk, H. J. 1981, \apj, 248, 344
%
\bibitem[Ellision et al.(1997)]{ellison97}Ellision, D. C., Drury, L. O'C., \& Meyer, J.-P., 1997, \apj, 487, 197
%
\bibitem[Ferrand \& Marcowith(2010)]{ferrand10}Ferrand, G., \& Marcowith, A., 2010, \aap, 510, A101
%
\bibitem[Gabici et al.(2009)]{gabici09}Gabici, S., Aharonian, F. A., \& Casanova, S., 2009, MNRAS, 369, 1629
%
\bibitem[Gargat\'{e} et al.(2010)]{gargate10}Gargat\'{e}, L., Fonseca, R. A., Niemiec, J., Pohl, M., Bingham, R., \& Silva, L. O., 2010, \apjl, 711, L127
%
\bibitem[Giacalone \& Jokipii(2007)]{giacalone07} Giacalone, J., \& Jokipii, J.R., 2007, \apjl, 663, L41
%
\bibitem[Gleeson \& Axford(1968)]{gleeson68}Gleeson, L. J., \& Axford, W. I., \apj, 154, 1011
%
\bibitem[Higdon et al.(1998)]{higdon98}Higdon, J. C., Lingenfelter, R. E., \& Ramaty, R., 1998, \apj, 509, L33
%
\bibitem[Inoue et al.(2009)]{inoue09} Inoue, T., Yamazaki, R. \& Inutsuka, S., 2009, \apj, 695, 825
%
\bibitem[Ioka (2010)]{ioka10}Ioka, K., 2010, Prog. Theor. Phys., 123, 743
%
\bibitem[Kashiyama et al.(2010)]{kashiyama10}Kashiyama, K., Ioka, K., \& Kawanaka, N., 2010, arXiv:1009.1141
%
\bibitem[Kawanaka et al.(2010)]{kawanaka10}Kawanaka, N., Ioka, K., \& Nojiri, M. N., 2010, \apj, 710, 958
%
\bibitem[Krymsky(1977)]{krymsky77} Krymsky, G. F., 1977, Doki. Akad. Nauk SSSR, 234, 1306
%
\bibitem[Li \& Chen(2010)]{li10} Li, H., \& Chen, Y.,2010, \mnras, 409, L35
%
\bibitem[Lingenfelter \& Higdon(2007)]{lingenfelter07} Lingenfelter, R. E., \& Higdon, J. C.,2007, \apj, 660, 330
%
\bibitem[Lucek \& Bell(2000)]{lucek00} Lucek, S. G., \& Bell, A. R. 2000, \mnras, 314, 65
%
\bibitem[Malkov \& V\"{o}lk(1995)]{malkov95}Malkov, M. A., \& V\"{o}lk, H. J, 1995, \aap, 300, 605
%
\bibitem[Malkov \& Drury(2001)]{malkov01}Malkov, M. A., \& Drury, L. O'C., 2001, Rep. Prog. Phys., 64, 429
%
\bibitem[Niemiec et al.(2008)]{niemiec08} Niemiec, J., Pohl, M., \& Nishikawa, K., 2008, \apj, 684, 1174
%
\bibitem[Ogliore et al.(2009)]{ogliore09}Ogliore, R. C. et al., \apj, 695, 666
%
\bibitem[Ohira et al.(2009a)]{ohira09a}Ohira, Y., Reville, B., Kirk, J. G., \& Takahara, F. 2009a, \apj, 698, 445
%
\bibitem[Ohira et al.(2009b)]{ohira09b}Ohira, Y., Terasawa, T. \& Takahara, F., 2009b, \apjl, 703, L59
%
\bibitem[Ohira \& Takahara(2010)]{ohira10}Ohira, Y., \& Takahara, F., 2010, \apjl, 721, L43
%
\bibitem[Ohira et al.(2010)]{ohira10}Ohira, Y., Murase, K. \& Yamazaki, R., 2010, \aap, 513, A17
%
\bibitem[Ohira et al.(2011)]{ohira11}Ohira, Y., Murase, K. \& Yamazaki, R., 2011, \mnras, 410, 1577
%
\bibitem[Riquelme \& Spitkovsky(2009)]{riquelme09} Riquelme, M. A. \& Spitkovsky, A., 2009, \apj, 694, 626
%
\bibitem[Panov et al.(2009)]{panov09}Panov, A. D. et al., 2009, Bulletin of the Russian Academy of Sciences: Physics, 73, 564
%
\bibitem[Parizot et al.(2004)]{parizot04}Parizot, E., Marcowith, A., van der Swaluw, E., Bykov, A. M., \& Tatischeff, V., 2004, \aap, 424, 747
%
\bibitem[Pato et al.(2010)]{pato10}Pato, M., Hooper, D., \& Simet, M., 2010, JCAP, 06, 022
%
\bibitem[Ptsuskin \& Zirakashvili(2003)]{ptuskin03}Ptuskin, V. S., \& Zirakashvili, V. N., 2003, \aap, 403, 1
%
\bibitem[Ptsuskin \& Zirakashvili(2005)]{ptuskin05}Ptuskin, V. S., \& Zirakashvili, V. N., 2005, \aap, 429, 755
%
\bibitem[Putze et al.(2010)]{putze10}Putze, A., Maurin, D., \& Donato, F. 2010, \aap, 526, A101
%
\bibitem[Smartt et al.(2009)]{smartt09}Smartt, S. J., Eldridge, J. J., Crockett, R. M., \& Maund, J. R. 2009, \mnras, 395, 1409
%
\bibitem[Strong et al.(2007)]{strong07}Strong, A. W., Moskalenko, I. V., \& Ptuskin, V. S., 2007, Annual Review of Nuclear and Particle Science, 57, 285
%
\bibitem[Tavani et al.(2010)]{tavani10} Tavani, M. et al., 2010, \apj, 710, L151
%
\end{thebibliography}
\end{document}